\documentclass[%
 reprint,
superscriptaddress,
 showkeys,
 nofootinbib,
 nobibnotes,
 amsmath,amssymb,
 aps,
 pra,
 floatfix,
]{revtex4-1}

\usepackage{graphicx}
\usepackage{dcolumn}
\usepackage{bm}
\usepackage{hyperref}
\usepackage{subfigure}

\usepackage[english]{babel}
\usepackage[utf8]{inputenc}
\usepackage[T1]{fontenc}
\usepackage{xspace}

\newcommand{\ls}{\ensuremath{\ell^\star}\xspace}
\newcommand{\ii}{\ensuremath{\mathrm{i}}\xspace}

\newcommand{\var}[1]{\ensuremath{\mathrm{Var}\left(#1\right)}}
\newcommand{\Var}[1]{\ensuremath{\mathrm{Var}\left(#1\right)}}

\usepackage{xcolor}

\begin{document}


\title{Magnetic field effects on one-dimensional Anderson localization of light}

\author{Lukas \surname{Schertel}}
\affiliation{%
 Fachbereich Physik, Universität Konstanz, 78457 Konstanz, Germany
}%
\affiliation{
 Physik-Institut, Universität Zürich, 8057 Zürich, Switzerland
}%
\author{Oliver \surname{Irtenkauf}}
\affiliation{%
 Fachbereich Physik, Universität Konstanz, 78457 Konstanz, Germany
}%
\author{Christof M. \surname{Aegerter}}
\affiliation{
 Physik-Institut, Universität Zürich, 8057 Zürich, Switzerland
}%
\author{Georg \surname{Maret}}
\affiliation{%
 Fachbereich Physik, Universität Konstanz, 78457 Konstanz, Germany
}%
\author{Geoffroy J. \surname{Aubry}}%
\email{geoffroy.aubry@unifr.ch}
\affiliation{%
 Fachbereich Physik, Universität Konstanz, 78457 Konstanz, Germany
}
\affiliation{%
 Departement de Physique, Université de Fribourg, 1700 Fribourg, Switzerland
}%

\date{\today}

\begin{abstract}
Transport of coherent waves in multiple-scattering media may exhibit fundamental, non intuitive phenomena such as halt of diffusion by disorder called Anderson localization.
For electromagnetic waves, this phenomenon was observed only in one and two dimensions so far. However, none of these experiments  studied the contribution of reciprocal paths nor their  manipulation by external fields.
In order to weaken the effect of reciprocity of coherent wave transport on Anderson localization 
in one dimension (1D), we studied light propagation through stacks of parallel Faraday-active glass slides exposed to magnetic fields up to 18 Tesla.
Measurements of light transmission statistics are presented and compared to 1D transfer-matrix simulations. The latter reveals a self-organization of the polarization states in this system leading to a saturation of the Faraday rotation-induced reciprocity breaking, an increase of the localization length, and a decrease of transmission fluctuations when reciprocity is broken.
This is confirmed experimentally for samples containing small numbers of slides while for larger samples a crossover from a 1D to a quasi-1D transport regime is found.
\end{abstract}

\maketitle


\section*{Introduction}

Anderson localization (AL) is a coherent wave transport phenomenon which results in the halt of diffusion due to interferences in multiple-scattering media.
Even though coherent light transport in highly multiple-scattering media has been studied for years \cite{Wiersma2013}, experiments which mainly focused on static or dynamic transmission properties still fail to show signs of this supposedly universal phenomenon for light in three-dimensional media \cite{Sperling2016,Skipetrov2016}.
While it is believed that this long-standing quest could be successfully achieved by further tuning  the sample properties by optimization of the scattering behavior~\cite{AubrySchertel2017,Schertel2019,Escalante2017},  signs of localization could be hidden behind other perturbing signals such as absorption or fluorescence \cite{Sperling2016}.
In other words, one may wonder whether the right quantities were measured up to now.
The quest for unequivocal signatures of AL is therefore of paramount importance~\cite{Smolka2011,Ghosh2017,Lopez-Bezanilla2018,Prat2019}.

The observation of the coherent backscattering cone~\cite{Wolf1985,*Albada1985} or weak localization, the precursor of AL, illustrated the importance of interferences in the description of randomly scattered wave phenomena.
Robust interferences  between reciprocal multiple-scattering paths ---which are at the heart of AL--- survive even in strongly disordered materials~\cite{Bromberg2016}.
Since coherent phenomena are very sensitive to the phase of the wave,
the magneto-optical Faraday effect which introduces circular birefringence between counter propagating waves~\cite{Rayleigh1885}
can be used to manipulate coherent wave  transport \cite{Golubentsev1984,MacKintosh1988}.

This effect breaks reciprocity  and therefore suppresses weak localization of light \cite {Erbacher1993,Lenke2000,Schertel2017}.
Moreover, a dephasing between counter-propagating paths was used to show the cancellinearly a fast step change in the refractive index of a random photonic medium~\cite{Muskens2012}, and the suppression and revival of coherent backscattering of ultra-cold atoms~\cite{Mueller2015} (as proposed by Ref.~\cite{Micklitz2015}).
These latter authors further suggested to break time-reversal symmetry to study AL of cold atoms.
This was done very recently by \citet{Hainaut2018} who used an artificial gauge field on cold atoms trapped in 1D to induce a parity-time symmetry breaking such that the system transits from the orthogonal to the unitary class.
Another  mechanism of controlled breaking of reciprocity of optical paths  by the Doppler effect in hot atomic vapor was discussed in~\cite{Cherroret2019}.    External magnetic fields also affect weak localization in electronic systems~\cite{Bergmann1984}.
In such systems, universal localization length changes have been predicted in quasi-1D (Q1D) systems when breaking reciprocity~\cite{Pichard1990,Thaha1993}, and non-universal changes in dimensions above 2~\cite{Lerner1995}.
Magnetic-field-induced delocalization was also reported in solid-state systems~\cite{Katsumoto1987,Lerner1995}, and several reviews \cite{Beenakker1997, Evers2008} address the topic of the impact of the external magnetic field on AL in electronic systems.

\begin{figure}
    \centering
    \includegraphics[page=1]{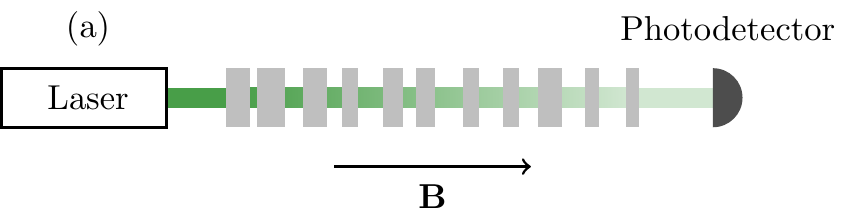}
    \includegraphics[page=3]{1D_setup.pdf}
    \includegraphics[page=4]{1D_setup.pdf}
    \caption{(a) Localization through a 1D stack of glass slides, and direction of the external magnetic field. (b),(c) Transition from 1D to quasi-1D due to spreading of the multiply reflected laser beam.}
    \label{fig:concept}
\end{figure}

Here we study the effect of magneto-optical dephasing on light localization in order to check its usefulness as an alternative probe of AL of light.
We test this idea in a system where light is known to localize, a 1D disordered system, as in 1D every infinitesimal amount of disorder is sufficient to localize the wave \cite{Abrahams1979,Berry1997}.
Note that this idea should not be confused with the magnetic field localization transition predicted in random ensemble of point scatterers~\cite{Skipetrov2015,*Skipetrov2018c}: in the latter system, the magnetic field does not act on reciprocity but on the interatomic dipole-dipole interaction that inhibits localization in such 3D systems without field~\cite{Skipetrov2014PRL,*Bellando2014}.
One-dimensional or quasi-one-dimensional electromagnetic wave localization can be realized 
either in confined geometries such as microwaves in a waveguide~\cite{Kuhl2008} or optical fibers~\cite{Karbasi2012}, or in open systems such as stacks of glass slides~\cite{Berry1997,Lu2010,Zhang2008,Park2010}, see Fig.~\ref{fig:concept}.
\citet{Bromberg2016} showed the effect of the manipulation of reciprocity on the coherent backscattering (weak localization) in a Q1D experiment using an optical fiber setup.
Here we use the Faraday effect to manipulate reciprocity in the {\em strong} localization regime  of stacks of Faraday active glass slides.

\citet{Berry1997} observed and explained why a stack of $N\gg 1$ transparent parallel plates of varying thickness acts essentially like a mirror: due to random interferences between waves multiply reflected backward and forward by the plates surfaces, the transmission $T$ of the intensity of the normal incident light decreases exponentially as a function of $N$,
\begin{equation}
\left<\ln T\right>=-2N \ln\left(\frac{1}{\tau}\right), \label{eq:exact}
\end{equation}
and light is mostly reflected.
This formula assumes that the wavelength of the light is much smaller than the deviations around the mean value of the plates thicknesses.
Here, $\tau=\frac{4n}{(n+1)^2}$ is the transmission of one glass$\rightarrow$air or air$\rightarrow$glass interface ($n$ is the refractive index of the glass) at normal incidence calculated by the Fresnel formulas, and the average $\left<\cdots\right>$ is done over the configurations. Note that this is different from the situation of incoherently added multiply reflected intensities which gives $T \propto 1/N$ (see, e.g., Ref.~\cite {Berry1997} and references there).
As the above expression includes reciprocal interferences, this situation was associated with localization by Berry and Klein~\cite{Berry1997}. The localization length $\xi$, corresponding to the inverse of the slope of Eq.~(\ref{eq:exact}), is given by  \begin{equation}
\frac{1}{\xi} =  2\ln\left(\frac{1}{\tau}\right) \label{eq:loclength}
\end{equation}
in units of plate number.
For stacks of glass plates in air  values of $\xi$ are typically $\leq 10$. Note that the exponential decay of Eq.~(\ref{eq:exact}) starts even for samples sizes $N$ shorter than the localization length $\xi$~\cite{Berry1997}.

The strong attenuation of the log-averaged transmission in the localized regime is also connected to a narrowing of the modes allowed to be transmitted.
This leads to enhanced fluctuations of the transmission. 
\citet{Chabanov2000} noted that the exponential decay of the static transmission with the sample thickness is not clearly distinguishable from absorption (similar to the 3D case) and that absorption may even suppress localization.
Moreover, the large fluctuations associated with localization imply that the mean transmission is not the natural scaling parameter. Rather, the full distribution or a parameter describing this distribution correctly is a better choice.
Thus, Genack and Chabanov~\cite{Genack2005} proposed to quantify the transmission fluctuations by using the \emph{variance} of the transmission $T$ \emph{relative} to its disorder ensemble averaged value $\langle T \rangle$ to observe localization in their samples, even in the presence of absorption.
\citet{Zhang2008} and \citet{Park2010}  then studied localization via thickness dependent transmission and transmission statistics, respectively, in a system showing a crossover from a 1D to a higher dimensional (>1D) geometry by sending visible light through a stack of not perfectly flat and not perfectly parallel glass slides.
Even though in this kind of system the transverse spreading of light is not bounded by reflecting boundaries, they showed that for a large number of slides it was described surprisingly well by the statistics of diffusive transport through long narrow tubes (Q1D) ~\cite{Kogan1995,Nieuwenhuizen1995}.
This is remarkable since the light  transport is rather different in both cases:  In the narrow tube geometry, the mean free path is isotropic and shorter than (or comparable to) the  tube diameter, while in the slide stack geometry, the mean free path is strongly anisotropic with a transverse value much larger than the beam diameter. Therefore, mode correlation effects are much stronger in the tube case. Nevertheless, for a small number of slides, one expects mode correlations in the stack system as well which implies that its transmission statistics should feature characteristics similar to perfect 1D localizing systems.
While such stacked systems are experimentally simple to realize, the Q1D geometry might weaken or even suppress localization for a large number of slides.

\begin{figure*}
    \includegraphics[width=\textwidth]{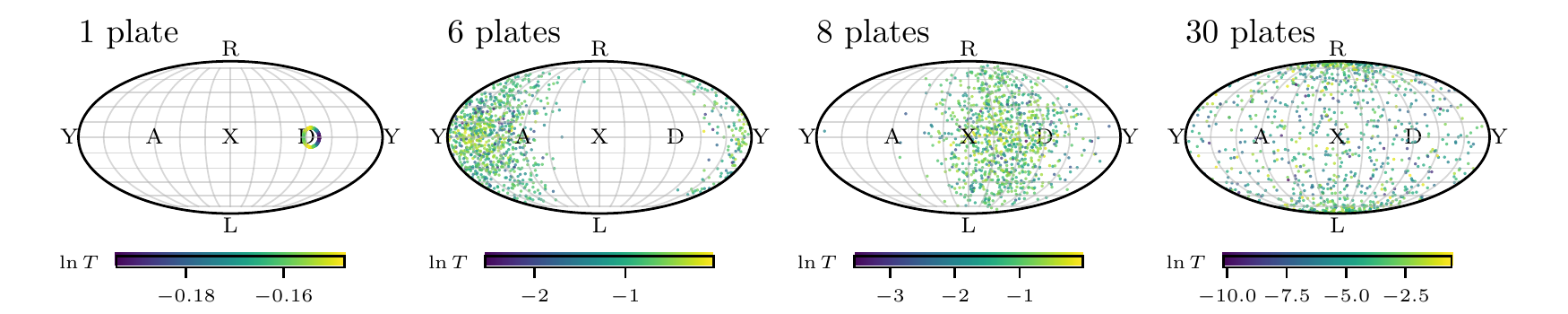}
    \caption{Simulated polarization states in transmission for samples of $N=1, 6, 8$ or 30 plates represented on the Poincaré sphere (Mollweide projection) at $B=18$\,T for the Faraday active plates studied in the paper.
The color encodes the logarithm of the total transmission ($T=T_x+T_y$).
X (respectively Y) represents linear polarization on the $x$ (respectively $y$) axis, D and A stand for diagonal and anti-diagonal (linear polarization on the first and second bisectors), and R/L are the circular polarizations (right/left).}
    \label{fig:poincareTrans}
\end{figure*}
In this paper, we study such a stack of slides made out of a Faraday active glass.
The interferences of the multiple reflections that lead to the exponential decay in transmission are studied by manipulating the phases of the waves via the Faraday effect in an external magnetic field applied parallel to the direction of propagation of light (see Fig.~\ref{fig:concept}(a)).
We compare our measurements with the results of transfer-matrix calculations from a perfect 1D stack of glasses with the same parameters.
The transfer-matrix calculations show an increase of the localization length and a decrease of the fluctuations in transmission when reciprocity is broken.
They further show an attraction of the polarization states towards the circular left and right polarizations, thus leading to a saturation of the reciprocity breaking mechanism in this 1D system.
In the experiments, the probability distribution of the normalized transmitted intensity allows us to identify the crossover range from 1D to Q1D due to imperfections in flatness and parallelism of the plates as sketched in Figs.~\ref{fig:concept}(b) and (c).
For samples smaller than this crossover length, the experiments show reasonable agreement with the 1D simulations and  represent the first experimental manipulation of optical AL via external fields.
For large sample thicknesses, high transmission fluctuations and AL are suppressed by the Q1D nature of the experimental sample.

The paper is organized as follows: First, in §~\ref{sec:numerical} we present the numerical results obtained within the transfer-matrix formalism for the propagation of an electromagnetic plane wave through an ideal stack of $N$ parallel Faraday active glass slides.
Next, in §~\ref{sec:experimentalconcept} the experimental concept is presented.
Finally, in §~\ref{sec:numericalandexp}, we compare the numerical 1D results with our experimental data.

\section{Numerical 1D results}
\label{sec:numerical}

In Appendix~\ref{sec:transfermatrix}, we present our extension of the transfer-matrix formalism in 1D systems in the presence of circular birefringence mechanisms such as Faraday rotation or optical activity.
The transfer matrices were calculated using Eq.~(\ref{eq:transferMatrix}), and the transmitted field was extracted using Eq.~(\ref{eq:solution}).
The parameters used for the simulations were chosen to be as close as possible to realistic experimental parameters. We therefore choose to work with $\lambda_0=532$\,nm (wavelength in vacuum), $n_0=1$ (refractive index of the air gaps), $n=1.8$ (refractive index of the Faraday active material at $B=0$\,T, Schott SF57~\cite{Goodman2015}), $V=31$\,rad/Tm (Verdet constant of the Faraday active material, Schott SF57~\cite{Goodman2015}), $d$ uniformly distributed on a span $2\Delta d=10\,\mu$m around $\left<d\right>=1.5$\,mm (thickness of both the Faraday active slabs and of the air gaps in between).
No absorption was put in the simulations.
We checked that the precision of the calculation was high enough to compute the coefficients of the large transfer matrices up to $125$ plates.
For each number of plates and for magnetic fields up to 25\,T, $3 \times 10^4$ different samples were calculated, and the whole statistics of all transmission and reflection coefficients was recorded.
With $a$ and $b$ being either $x$ or $y$, the transmission (respectively reflection) coefficients are defined as follows: $T_{ab}=\left|E_{N}^{b\rightarrow}/E_{0}^{a\rightarrow}\right|^2$ (respectively $R_{ab}=\left|E_{0}^{b\leftarrow}/E_{0}^{a\rightarrow}\right|^2$) are the transmitted (respectively reflected) intensities with linear polarization $b$ from incident linear polarization $a$, $T_a=T_{ax}+T_{ay}$ is the total transmitted intensity for incident linear polarization $a$.
At $B=0$\,T, the two polarization channels are uncoupled, and therefore $T_{xy}=T_{yx}=0$.
Moreover for symmetry reasons, at any field, $T_{xx}=T_{yy}$, $T_{xy}=T_{yx}$, and therefore $T_x=T_y$.

In our 1D model, one major difference between optical activity and Faraday rotation is that in the first case, the rotation angle of the polarization plane of each photon is proportional to the \emph{position in the sample}, while in the latter it is proportional to the \emph{path length of the photon in the sample}.
We therefore expect all reflected photons in the optical activity case to come back with the same polarization as the incident photons, and to mix the polarizations in the Faraday rotation case.
This is indeed observed in the numerical simulations (see Appendix~\ref{sec:polStates}).
Looking at how the reflected intensities distribute over the $x$ and $y$ polarization channels for both optical activity and Faraday rotation as a function of the sample length (see Appendix~\ref{sec:reflectionCoeffFROA}) further illustrates the fact that Faraday rotation breaks reciprocity, while optical activity does not.

In transmission we also observe a spreading of the polarization states  as shown in Fig.~\ref{fig:poincareTrans}.
The polarization states for 1000 different samples consisting of 1, 6, 8 or 30 plates are represented on the Poincaré sphere which represents the Stockes parameter of polarized light.
The important polarization states are described in the legend of Fig.~\ref{fig:poincareTrans}.
Without magnetic field, all the points are on the X point: light stays linearly polarized.
Applying a magnetic field induces a rotation of the points on the equator (linearly polarized light) but also a spreading out of the equator (elliptically polarized light).
When the number of plates and the external field are high enough the points are attracted by the poles (as observed for $N=30$ plates in Fig.~\ref{fig:poincareTrans}).
The poles are the circularly polarized states.  Those circularly polarized states are not sensitive to the field induced circular birefringence and therefore do not allow for magnetic-field-induced reciprocity breaking: While Faraday rotation changes the phase of a path with a given circular polarization, this phase change is identical to the one of the reciprocal path with the same polarization, such that there is no field dependent relative phase shift between both paths. Therefore, as soon as points on the Poincaré sphere are getting close to the poles, they diffuse less: the effect of the external magnetic field decreases.
Note that this system therefore shows a manifestation of self-organization of the polarization state in an optical system~\cite{Guasoni2016,Vorontsov1995}.

\begin{figure}
    \includegraphics[width=\columnwidth]{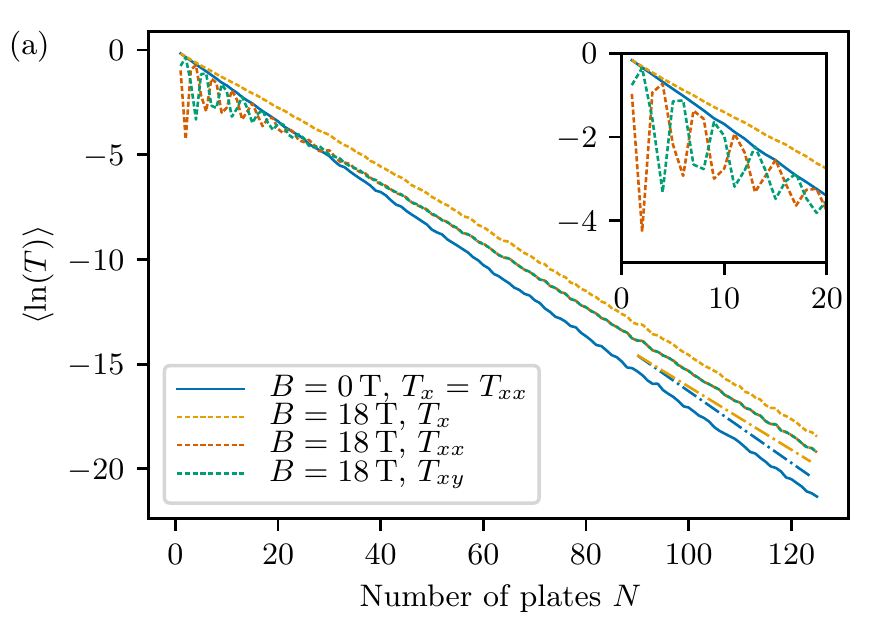}
    \includegraphics[width=\columnwidth]{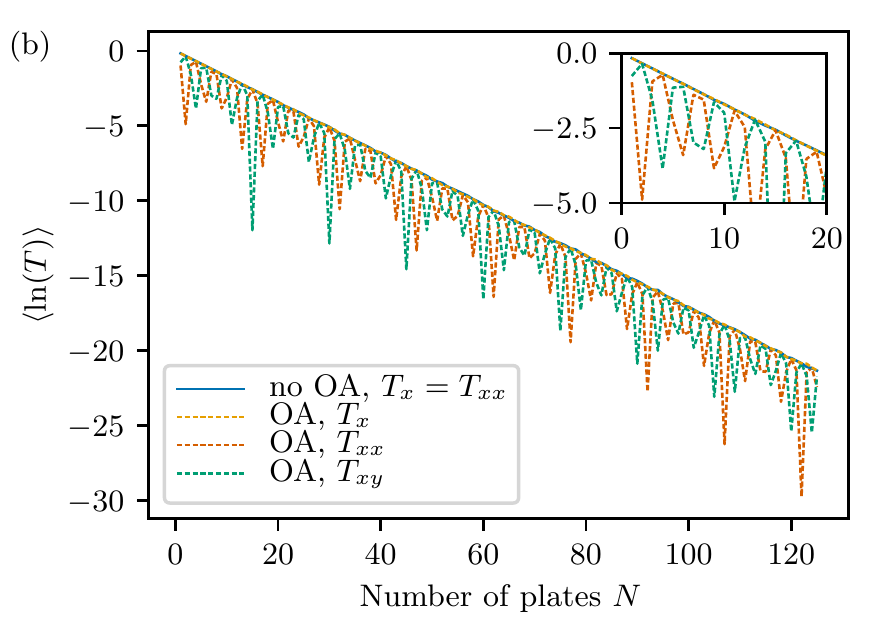}
    \caption{Average of the logarithm of $T_x$, $T_{xx}$ and $T_{xy}$ without and with (a) Faraday rotation or (b) optical activity for the same rotation angle of 48$^\circ$ for a single passage through a single plate. This corresponds to a circular birefringence $\Delta n=4.72\times 10^{-5}$ for FR in SF57 at $B=18$\,T and at the average thickness of the plates.}
    \label{fig:lnT}
\end{figure}

In Fig.~\ref{fig:lnT}(a), we plot the different transmission coefficients for the intensity at 0 and 18\,T when Faraday rotation is present.
At $B=0$\,T, $\left<\ln(T_{x})\right>=\left<\ln(T_{xx})\right>$ decreases linearly with the number of plates as expected (see Eq.~(\ref{eq:exact})) \cite{Berry1997}. 
The extracted localization length is $\xi=5.85$ plates, which is very comparable to the 5.87 plates value obtained from Eq.~(\ref{eq:loclength}). This clearly shows that 1D Anderson localization occurs in this system.
However, the curves are qualitatively different when the external magnetic field is turned on.
At $B=18$\,T, oscillations are observed in both $\left<\ln(T_{xx})\right>$ and $\left<\ln(T_{xy})\right>$ up to about 30 plates.
These oscillations come from the fact that for a small number of plates, the proportion of light transmitted without reflection is high, 
such that the mean polarization state is governed by the Faraday rotation of the light through the stack in one direction.
Looking for instance only at $\left<\ln(T_{xx})\right>$ leads to a diminished transmission for 2, 6, ... plates where the mean polarization is perpendicular to the incident polarization.
Calculating the Faraday rotation of the directly transmitted light as a function of the total length of Faraday material in the sample, one expects a perpendicular polarization for 1.9 plates, 5.6 plates, etc., consistent with the latter observation. At these specific numbers, mainly multiply reflected light (which has undergone some more or less stochastic polarization changes as shown in Fig.~\ref{fig:poincareTrans}) is detected, which gives rise to an increased observation of localized light and thus a decreased $\left<\ln(T_{xx})\right>$.
For an increasing number of plates, more multiply reflected photons govern the mean transmission, such that the distribution of polarization gets further broadened.
For $N\gtrsim 30$, $T_{xx}=T_{xy}$, and therefore $\left<\ln(T_{x})\right>=\left<\ln(T_{xx})\right>+\ln 2$. In this regime, the exponential decrease of the mean transmission is slowed down with magnetic field.
The ratio of the slopes of the linear fits to the $B=0$\,T and $B=18$\,T data is found to be $r=1.1130\pm0.0009$.
This suggests that the localization length in the system is increased by a factor $r$ when the magnetic field is on.
The subsistence of an exponential decay with magnetic field may suggest that the magnetic field only disturbs some of the localized modes. In fact, however, for sufficiently high values of $N$ (at high enough $B$) the mixed elliptical polarization states progressively evolve into equal amounts of the two circular polarized states (as illustrated above) which cannot evolve further because the latter do not couple to FR. Therefore AL modes cannot be manipulated further beyond this point. Increasing the product of the magnetic field times the Verdet constant decreases the quasi-period of the oscillations and moves the onset of the circular polarized regime to a smaller number of plates.

We also ran the calculation for different values of the refractive index $n$ of the Faraday active material without any external magnetic field.
Without magnetic field, we verified that the slope of $\left<\ln(T_{xx})\right>$ as a function of $N$ is equal to $1/\ln \tau$ (Eq.~(\ref{eq:exact}), \cite{Berry1997}).
Even with this dependency, in the range explored ($1.4<n<2.0$, $B=9, 18, 27$\,T), we find that $r$ neither depends significantly on $n$ nor on $B$, once $B$ is high enough ($r=1.113 \pm 0.003$, mean and standard deviation on all the calculated data).
This result seems similar to the universal relation between the localization length with and without time-reversal symmetry found for Q1D electronic~\cite{Pichard1990} and quantum chaotic systems~\cite{Thaha1993} with a large number of transverse modes, whereas the case studied here is purely one-dimensional, that means with only one single transverse mode.
However, the relation found for the latter systems (removing completely time-reversal symmetry doubles the localization length) does not seem to hold in our case.
One reason could be that time-reversal symmetry is not completely broken in our case: even though increasing the magnetic field does not alter the results anymore, the attraction of the polarization states toward the circular polarized states for long samples (see Fig.~\ref{fig:poincareTrans}) cancels time symmetry breaking.
Another reason could also be the fact that our system has two polarization modes that are either not coupled (without field), or field-dependent mixed with field.
This is different from the $\mathcal{N}$ perfectly mixed modes of the Q1D literature.
In Ref.~\cite{Beenakker1997}, Eq.~(220) states that for $\mathcal{N}=1$ mode (pure 1D system), the localization length should be independent of the magnetic field ($r=1$), whereas for $\mathcal{N}=2$ modes (two coupled polarization states), $r=4/3$.
Our results are consistent with neither predictions, but the $\sim 10\%$ change is clearly established in our numerical results.
To illustrate this clear change,  the same analysis is done for optical activity (OA) in Fig.~\ref{fig:lnT}(b).
In this case, no change in the localization length is observed.
Note that for the OA case, oscillations with increasing number of plates are preserved in contrast to the Faraday rotation case.
For OA, the polarization state in transmission only depends on the sample size in which case relative fluctuations from sample to sample vanish for large $N$.

\begin{figure}
    \includegraphics[width=\columnwidth]{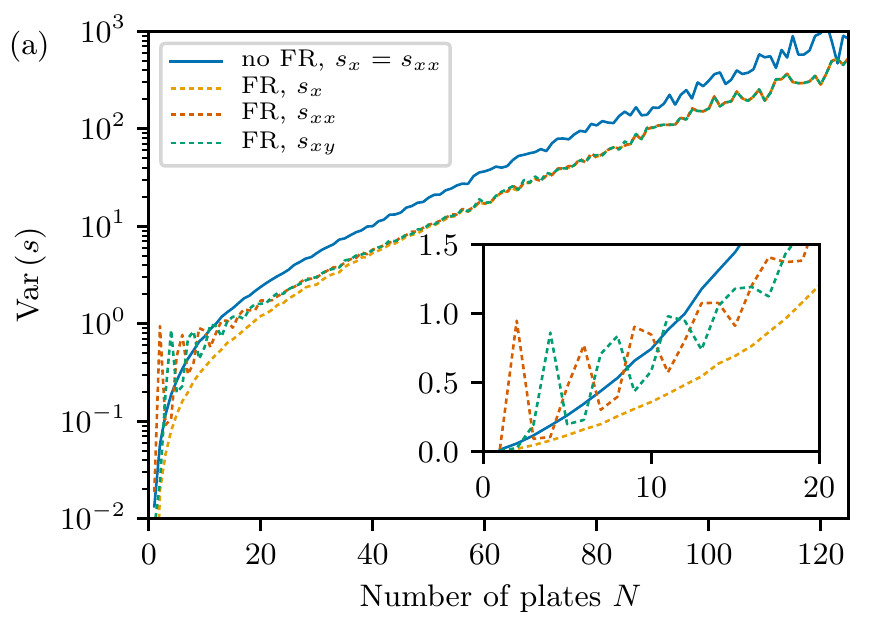}
    \includegraphics[width=\columnwidth]{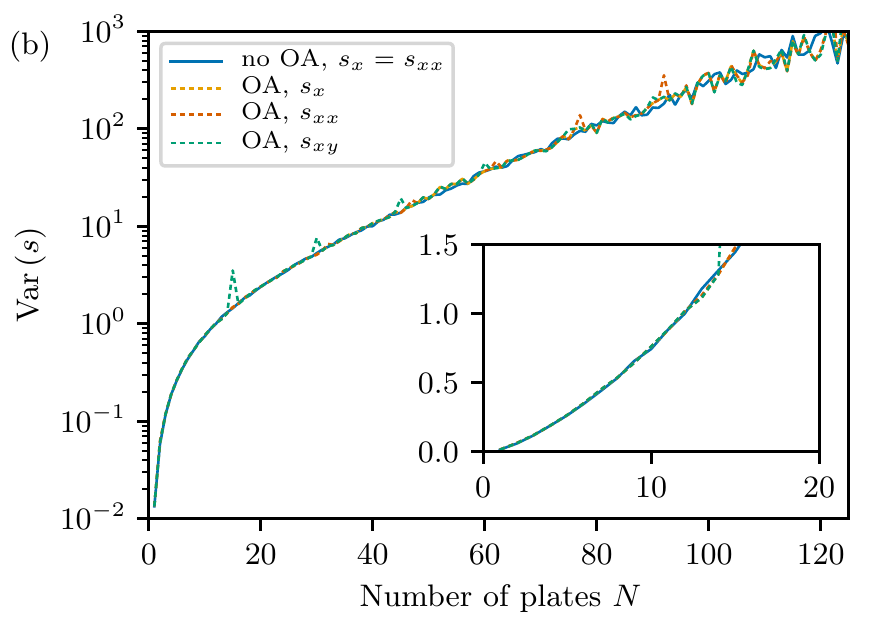}
    \caption{Variance of the normalized transmitted intensity $\Var{s}$ as a function of the number of plates $N$.
        (a) Faraday rotation, and (b) optical activity.
        Similarly to Fig.~\ref{fig:lnT}, the birefringence and the average thickness of the plates are the same for the FR and OA simulations.
    }
    \label{fig:var}
\end{figure}

In 1D systems, the variance of the normalized transmitted intensity $\var{s}$, with $s=T/\left<T \right>$ is expected to increase exponentially with number of plates $N$ indicating a strong increase of the fluctuations in transmission in the localized regime~\cite{Chabanov2000}.
The blue solid line in Fig.~\ref{fig:var}(a) shows such an increase without magnetic field (no FR).
With a magnetic field ($B=18~$T) such an exponential increase is still present for all normalized transmissions $s_x=T_x/\left<T_x \right>$, $s_{xx}=T_{xx}/\left<T_{xx} \right>$ and $s_{xy}=T_{xy}/\left<T_{xy} \right>$, but on a lower scale. This indicates that high transmission fluctuations present in the localized regime are partly suppressed by the Faraday effect. This observation is consistent with the observation of an increased localization length in Fig.~\ref{fig:lnT}(a). Again oscillations are observed in both $\var{s_{xx}}$ and $\var{s_{xy}}$ for a small number of plates (see inset of Fig.~\ref{fig:var}(a)). The selective detection of multiply reflected light for a certain number of plates by polarization selection increases the transmission fluctuations and leads to enhanced signs of localization similar to Fig.~\ref{fig:lnT}(a). For $N>30$ the distribution of polarization states is {further} broadened such that $\var{s_{x}}=\var{s_{xx}}=\var{s_{xy}}$.

In Fig.~\ref{fig:var}(b), the same analysis is performed for optical activity (with equal rotation of the polarization) instead of Faraday rotation.
In this case, no change in the statistics is observed.

\section{Experimental concept}
\label{sec:experimentalconcept}

\begin{figure}
    \centering
    \includegraphics[width=0.5\columnwidth]{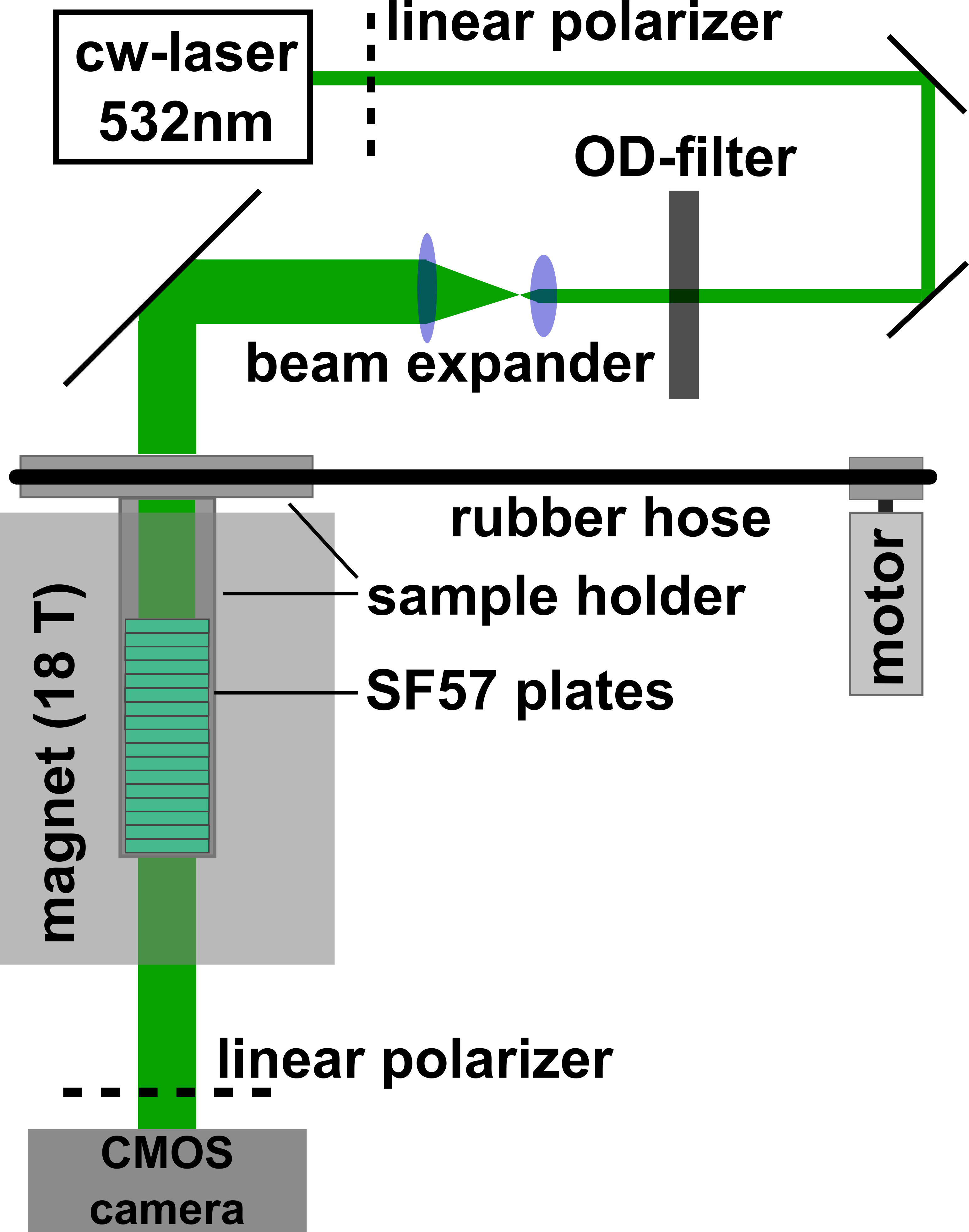}
    \caption[1DSetup]{1D light transport setup: a laser beam is sent through an OD filter to control the incident intensity before being expanded to $\approx 0.5$\,cm. This expanded beam illuminates the sample made of SF57 plates (varying from 0 to 50) placed inside the 18\,T superconducting magnet.
    A linear polarizer can be placed in front of the CMOS camera to select one plane of polarization.
    The sample is placed inside a sample holder which is rotated constantly via a rubber hose by a motor.}
    \label{1DSetup}
\end{figure}

We describe now our experiment to study effects of reciprocity on Anderson localization of light.
We measured transmission statistics through a stack of Faraday active slides as a function of an external applied DC magnetic field.
We designed the experimental setup (see Fig.~\ref{1DSetup}) such that the linear incoming polarization was Faraday-rotated by about $ 45 ^{\circ}$ for propagation through one plate under the maximum available magnetic field (18\,T) which was produced by a superconducting solenoid with 41-mm-diameter vertical room-temperature bore (Oxford Instruments).
This FR was chosen to   completely randomize the phases of the waves on their way through the stack.
We therefore used round slides (2.5\,cm diameter, $\left<d\right>\approx 1.5$~mm) of Schott SF57 glass, a dense flint glass (high refractive index $n=1.8$ caused by the high lead oxide concentration in the material~\cite{SchottSF57}) with strong Faraday rotation~\cite{Weber2002}.
The Verdet constant was measured to be $V\approx31.0 \pm 0.1~\mathrm{rad}/\mathrm{Tm}$ (in good agreement with~\cite{Weber2002}) which corresponds to a rotation angle of $\theta\approx48 ^{\circ}$ for one plate at $\lambda_0=532$\,nm and $B=18$\,T. 

Putting the plates on top of each other leads to somewhat irregular spacings between the plates and slight relative inclinations of their normals (see Fig.~\ref{fig:concept}(c) and Refs.~\cite{Zhang2008,Park2010}). This is because the plates were not specially treated to be optically flat so that they have a surface roughness comparable to the wavelength of the incident light. In addition, their  thickness may not be perfectly equal nor perfectly homogeneous and their normals rather slightly tilted off the direction of the incident beam. This leads to varying spacings and thicknesses at different transverse  positions of the stack.
The idea of this setup is to send a relatively wide ($\approx 0.5$~cm) parallel coherent light beam through this disordered stack of $N$ glass plates ($0 < N < 50$) and to monitor the transmission patterns in the far field by a camera.
On progression through the stack the phase of the incident plane wave obeys different shifts at different lateral positions and the wave fronts become more and more distorted with increasing $N$ as further discussed in the following section.

The sample holder was continuously rotated about an axis normal to the stack via a rubber hose by a motor.
The motor turns the sample by one revolution  per minute which corresponds to the measurement time for each sample.
During this time the camera (a 12-bit CMOS camera with 1900$\times$1200 pixels from The Imaging Source, DMK 23UX174) took 300 images (5 per second). The exposure time was chosen to be small enough ($10^{-4}$\,s) to minimize the averaging which occurs by turning the sample.
The linearly polarized coherent laser beam comes from a continuous laser (Torus, Laser Quantum, coherence length $>100$\,m). The output power can be tuned between $P=0.5$\, W and 0.75\,W. An optical density (OD) filter is used to further control the incident intensity to avoid camera saturation.
The beam is expanded using two lenses with $f_1=200$\,mm and $f_2=1000$\,mm in a confocal setting, and then led through the sample in the magnet.
The beam was adjusted to pass parallel through the relatively long (1174\,mm) bore by aligning the back reflection of the first plate back on the incident beam. 
The measurements involving $T_{xx}$ were done with a linear polarizer in front of the camera to capture only the photons with the same polarization as the incident light, and $T_{x}$ was measured without a polarizer.

\section{Experimental results vs 1D numerical simulations}
\label{sec:numericalandexp}

\begin{figure}
    \includegraphics[width=\columnwidth]{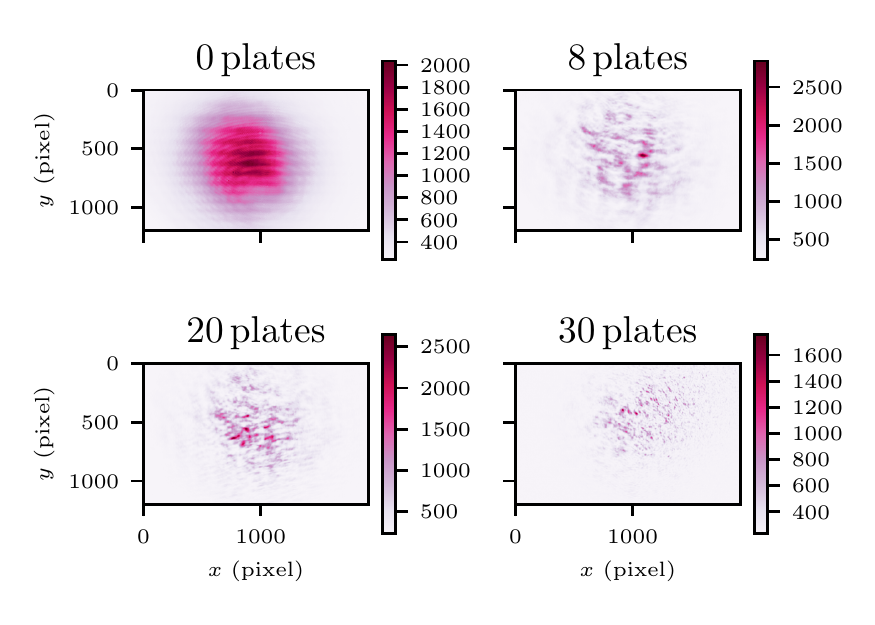}
    \caption{Measured images for four increasing numbers of plates ($N=0$, $5$, $15$, $20$). The measurement was performed at $B=0T$ without polarizer in front of the camera. Color bars show measured intensities in a.u.}
    \label{fig:pic-plates}
\end{figure}

Typical snapshots of the evolution of the transmission patterns with increasing number $N$ of slides in the stack are shown in Fig.~\ref{fig:pic-plates}.
\begin{figure*}
    \includegraphics[width=\textwidth]{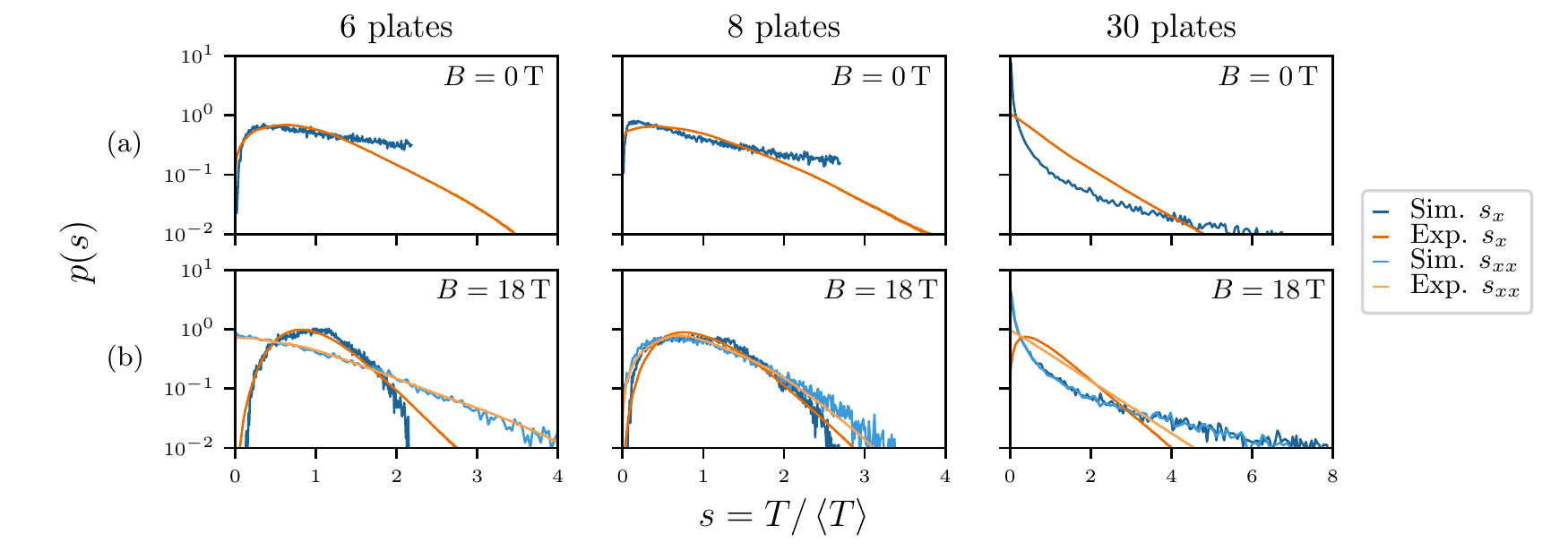}
    \caption{Histograms of the normalized transmitted intensity $s=T/\left<T\right>$ without (upper row (a)) and with (lower row (b), $B=18$\,T) external magnetic field for $N=6, 8$ and 30 plates.
    Simulation data are traced in blue and experimental data in orange.
    }
    \label{fig:hist}
\end{figure*}
For $N=0$ (no sample) the intensity profile across the image surface is not perfectly Gaussian as would be expected for a Gaussian incident beam  but some low contrast interference fringes can be noticed. They are due to the entrance window of the CMOS camera which is somewhat inclined with respect to the optical axis.
With increasing $N$, an increasing number of similar fringe patterns of slightly inclined sample slides coherently superimpose in the transmitted pattern which, therefore, becomes more fine grained with increasing length of the stack.  For $N=8$ plates some residual fringes can still be seen mixed with a more irregular pattern. The image with 20 plates shows that an almost random pattern  modulates the beam profile and  the intensity pattern becomes close to the familiar random interference pattern (speckle) characteristic for fully randomly distributed phases \cite{Goodman2015}. This is because with increasing $N$ the wavefront distortions rapidly become larger than the wavelength and the patterns become more irregular because of the increasing contributions of long multiply reflected waves. These observations are qualitatively similar to those reported by \citet{Zhang2008}. 

\subsection{Zero-field transmission statistics}

The non ideal alignment of the glass plates and their surface roughness thus result in a crossover from a situation of interfering plane waves at small $N$ to a Q1D scattering geometry for a large $N$ similar to the experimental realization by \citet{Park2010}. In the following section we use the comparison between experimental and numerical intensity statistics to check how close our experimental system comes to the simulated  ideal 1D case, with an emphasis on experimental signatures of 1D AL.  We anticipate that the Q1D case at high $N$ can not be captured by the numerical model.

Figure~\ref{fig:hist} displays the probability distributions of the transmission $p(s)$, with $s=T/\langle T \rangle$, for 6, 8 and 30 plates, the upper (a) and lower (b) row corresponding to zero and maximum field, respectively.
Experimental data are shown in orange while the 1D simulations are plotted in blue. Overall, we find reasonable agreement between experimental and simulated intensity distributions for low $N$, in particular at high magnetic fields, while differences become progressively more pronounced with increasing $N$.

We consider first the zero-field case (a) in more detail. For small $N$ (6 or 8), both curves show a similar functional form at small and intermediate $s$ with a low value for $p(s)$ at small $s$ followed by a maximum. This compares well with the 1D simulations of \citet{Park2010} and shows that our experimental results are close to the 1D situation at low $N$.
Experimental $p(s)$ deviate, however, substantially from our simulations at high intensities in lacking the algebraic high intensity tail (see the $N=30$ plates case in Fig.~\ref{fig:hist}(a)). We attribute this essentially to field correlations visible by the stripe  interferences described above which particularly spoil the brightest localized modes requiring constructive interferences of the longest multiple reflected paths. In addition, we believe that in our experimental setting rotating the sample does not necessarily provide a good ensemble average over a sufficient number of independent disordered configurations as the fringes are seen to rotate with the sample. For increasing $N$, the simulations exhibit a more and more pronounced algebraic shape with increasing contributions at low and high intensities, which highlights the anomalously high transmission fluctuations characteristic for localization.
The experimental distributions, however, are severely affected by the crossover from the experimental 1D fringe situation to the Q1D speckle situation~\cite{Park2010} and therefore tend towards an exponential decay typical for a classical polarized speckle (see the $N=30$ plates case in Fig.~\ref{fig:hist}(a)). Note  that, consistently, by comparing data taken with and without an analyzer we verified that the polarization stays linear up the highest $N$ for the case $B=0$\,T.

\subsection{High-field transmission statistics}

The lower row (b) of Fig.~\ref{fig:hist} shows that with a magnetic field ($B=18$~T), the probability distributions of the normalized transmission $s$  obey dramatic changes for both  simulations and experiments.
For small $N$ (here we show data for $N=6$ and $N=8$) we find excellent agreement between simulations and experiments.
In the case of no polarization selected detection ($p(s_x)$) the distributions are relatively narrow and similar to those of a superposition of two orthogonal polarized speckle intensities \cite{Goodman2015}. This is due to the strong and widely distributed Faraday rotations which result in a strong depolarization already after a few plates as illustrated in Fig.~\ref{fig:poincareTrans} and in Appendix~\ref{sec:polSpreadingExp}. The polarization selected measurements ($s_{xx}$) perfectly agree with the simulations for $N=6$ and $N=8$. This is because the strong field induced path-length dependent depolarization generates enhanced phase fluctuations for any given polarization which leads to a lower fringe contrast and to a much better configuration averaging as in the zero-field case. At high $N$ the experimental $p(s)$ tend towards the distributions of a classical polarized ($p(s_{xx})$) and fully depolarized ($p(s_x)$) speckle, respectively, as expected for Q1D (see the $N=30$ plates case in Fig.~\ref{fig:hist}(a)). Therefore, the deviations between experiment and simulations at high $N$ are due to the 1D to Q1D crossover in the experiment which spoils signatures of localization.

\begin{figure}
    \includegraphics[width=\columnwidth]{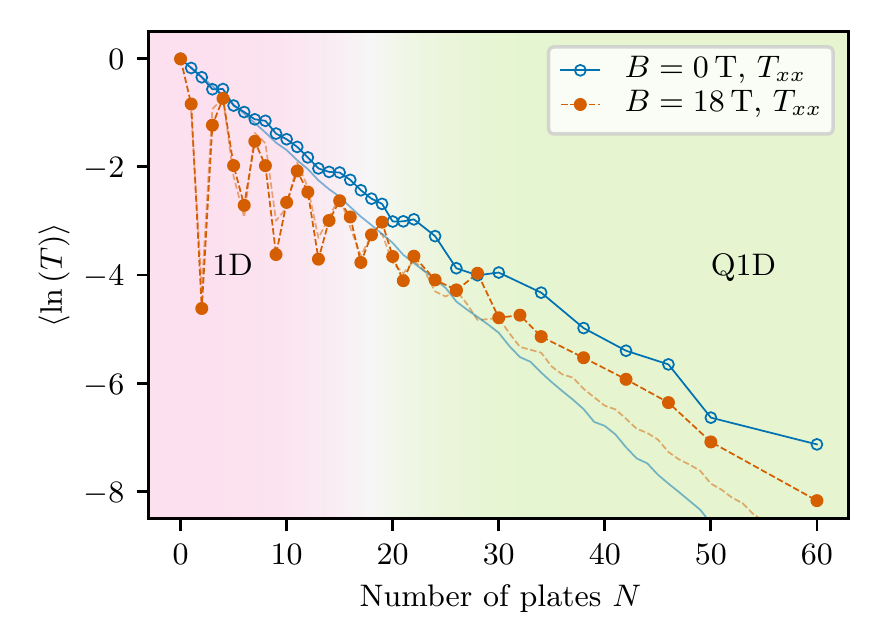}
    \caption{Experimental data of the mean logarithmic relative transmission as a function of the number of plates $N$ at $B=0$\,T and 18\,T.
    Data are measured with a polarizer at the output.
    The light curves are the simulation results (same data as in Fig.~\ref{fig:lnT}(a)).}
    \label{fig:N_StaticTransExp}
\end{figure}

The difference between the functional forms of $p(s_{xx})$ for $N=6$ and $N=8$ also reflects the oscillations in $\langle \ln \left( T_{xx} \right) \rangle $ in the simulations shown in Fig.~\ref{fig:lnT}(a).
Such oscillations are also present in the experiments: in Fig.~\ref{fig:N_StaticTransExp}, we plot $\left<\ln T_{xx}\right>$ as a function of the number of plates without and with magnetic field.
Without external magnetic field (blue open circles in Fig.~\ref{fig:N_StaticTransExp}), $\left<\ln T_{xx}\right>$ is expected to decrease linearly~\cite{Berry1997} with $N$ as also observed in the transfer-matrix calculations (blue line in Fig.~\ref{fig:lnT}(a) or light blue line in Fig.~\ref{fig:N_StaticTransExp}).
This is well verified for low numbers of plates ($N<10$) and deviations to this linear behavior due to the Q1D situations are then observed.
When the magnetic field is turned on, the oscillations of $\left<\ln T_{xx}\right>$ (orange solid circles in Fig.~\ref{fig:N_StaticTransExp}) due to the contribution of the non-reflected light are seen with the same period as in the 1D simulations (dark orange dashed line in Fig.~\ref{fig:lnT}(a)).
In the transfer-matrix calculations the slope of the decay with $N$ is smaller for the measurement with magnetic field than without.
This was connected to an increased localization length $\xi$.
However, in the experiment no such change of the slope is observed. This is
explained by the fact that in the regime where this change is expected to occur ($N>20$), signs of localization are already affected by the Q1D nature of the scattering system.

The change of the probability distribution of the normalized transmission with magnetic field (6 plates and 8 plates in Fig.~\ref{fig:hist}) as well as the oscillations in $\langle \ln T_{xx} \rangle$ with magnetic field, both in the experiments and in the simulations, are signs of manipulation of AL via magnetic fields in a perfect 1D geometry where localization exists from the first plate on.
Even though the samples are short compared to the mean-free path, 1D systems are always localized ---even for very small samples--- as the exponential decay starts from the first plate on (see Fig.~\ref{fig:lnT}(a)).
A further probe on manipulation of reciprocity and thus AL by magnetic fields via the Faraday effect will be done in the following by analyzing the variance of the normalized transmitted intensity $\var{s}$.

\subsection{Effect of magnetic field on Anderson localization}

\begin{figure}
	\includegraphics[width=\columnwidth]{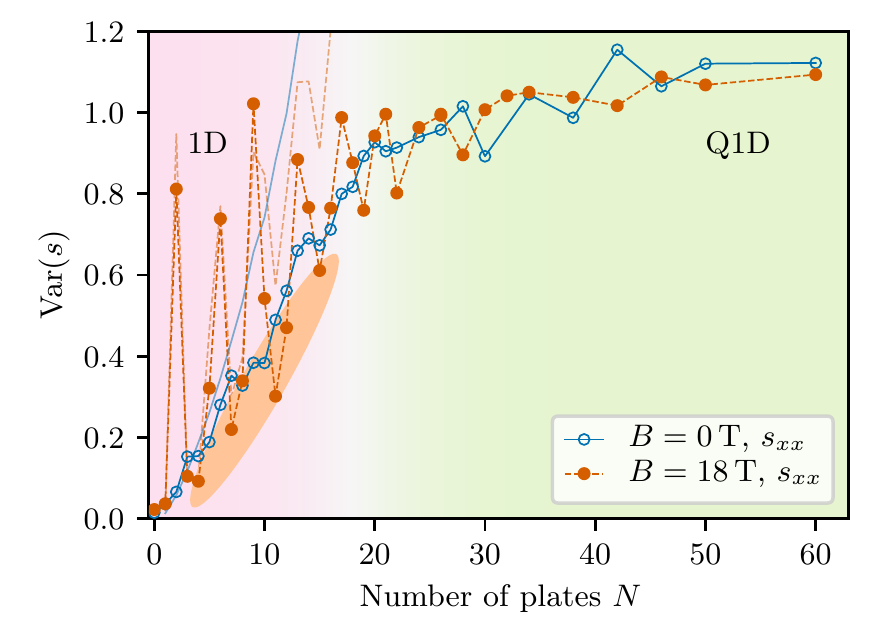}
    \caption[VarT]{Experimental data of the variance of the normalized transmitted intensity $\var{s}$ through a stack of $N$ Faraday active glass slides with (orange) and without (blue) magnetic field.
    Data are measured with a polarizer at the output.
    The ellipse highlights the measurements where $\var{s_{xx}}$ is lower with magnetic field than without in the 1D regime.
    The light curves are the simulation results (same data as in Fig.~\ref{fig:var}(a)).}
    \label{fig:VarToverN}
\end{figure}

The main purpose of this paper is to study the effect of breaking reciprocity on Anderson localization via Faraday rotation.
Figure~\ref{fig:VarToverN} shows the experimental data for $\var{s}$ versus $N$ without (blue open circles) and with (orange filled circles) magnetic field. For a small number of plates ($N<8$), a similar
increase as in the simulations is observed.
In this regime, oscillations are again observed in the magnetic-field measurements.
Similar to the simulations (inset in Fig.~\ref{fig:var}(a)) the minima of these oscillations fall below the $B=0$~T measurement (see orange ellipse in Fig.~\ref{fig:VarToverN}).
This indicates the early stage of a slower increase of $\Var{s_{xx}}$ and thus confirms the increase of the localization length with magnetic field found in the simulations.
At $N>20$ plates for all measurements with and without magnetic field a plateau occurs.
The occurrence of such a plateau in the variance of the transmitted intensity was interpreted by \citet{Park2010} as a signature of the crossover from 1D to Q1D.
High intensity fluctuations are suppressed by mode coupling in Q1D geometry as observed in Fig.~\ref{fig:hist} for $N=30$.
The crossover range is between $N=10$ and $N=20$, consistent with the probability distributions of Fig.~\ref{fig:hist}. This happens to be close to the expected localization length $\xi\approx 6$\,plates, such that no difference in the plateau value is observed without and with magnetic field $B$ for $s_{xx}$. 
The relatively thick plates (necessary to obtain a large rotation angle of Faraday rotation) leads to an earlier crossover from 1D to Q1D (\citet{Park2010} used plates of average thickness 150~$\mu$m) and suppress localization for large $N$.
Note that on the other hand the high refractive index of SF57 ($n=1.84$ at $\lambda_0=633$~nm~\cite{Weber2002}) lowers the localization length $\xi$ by increasing the scattering (i.e., reflection) compared to Park \textit{et al.} experiments($n=1.5$ \cite{Park2010}).

\section*{Conclusion}
We used transfer-matrix simulations to show the effect of breaking reciprocity on 1D Anderson localization of light via the magneto-optic Faraday effect. The simulations show a magnetic-field-induced increase of the localization length by a factor which neither seems to depend on the refractive index of the plates nor on the magnetic field once Faraday rotation is high enough.
This is associated with a  decrease of the transmission fluctuations which was also observed in the simulations and confirmed in experiments.
The non-trivial change in the localization length is related to the fact that for high $N$ and $B$, the polarization states are spread, but as soon as they come close to the circular ones, they do not couple anymore to the external magnetic field because reciprocity is not broken for circular polarization in a Faraday active medium.
The experimental data are a clear observation of manipulation of AL of light by external magnetic fields. For a small number of plates, the experimental data showed a similar behavior as predicted by the numerical simulations. For stacks between 10 and 20 plates a crossover from 1D to Q1D is observed.
To be able to observe localization in 1D experiments even for large $N$ the crossover length from 1D to Q1D would need to be increased. This might be achievable by a better parallel  alignment of the glass plates in the stacks, e.g., by using  piezo elements between plates, or by thinner glass plates keeping the FR high enough to randomize the polarization state. The first solution leads to a very sophisticated experimental sample, which might be hard to realize. The second solution may be easier to realize.

In conclusion, we show in this paper that Anderson localization of light is indeed affected by reciprocity breaking, even though the 1D configuration studied here does not allow it to be fully broken.
This paves the way towards a new probe for Anderson localization of light in higher dimensional systems.

\begin{acknowledgments}
We thank Luis S. Froufe-Pérez for fruitful discussions.
We acknowledge support by the Center for Applied Photonics (CAP, project CAP05), University of Konstanz, and the Schweizerischer Nationalfonds (SNF, Grant No. 200020M\_162846).
\end{acknowledgments}

\clearpage

\appendix
\section{Transfer matrix}
\label{sec:transfermatrix}

\subsection{1D Transfer-matrix formalism and Faraday rotation}
In this section, we introduce the transfer-matrix formalism in the presence of Faraday rotation.
To make our paper self-contained, we start from the usual definition of the transfer matrices.
\begin{figure}[b]
    \centering
    \includegraphics{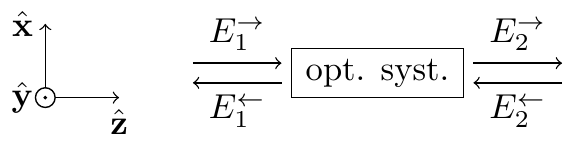}
    \caption{Direction of propagation of the light electromagnetic fields $E$ in front (1) and behind (2) any optical system (dielectrics, interface,~...).}
    \label{fig:1Dgeometry}
\end{figure}
Given any optical system (see Fig.~\ref{fig:1Dgeometry} for the geometry), the transfer-matrix formalism decomposes the field in plane waves propagating in the forward $+z$ ($\rightarrow$) and backward $-z$ ($\leftarrow$) directions, and relates the field behind the system to the field in front of the system through the transfer matrix $\mathcal{M}$,
\begin{equation}
\begin{pmatrix}E_2^\rightarrow \\ E_2^\leftarrow\end{pmatrix} = \mathcal{M} \begin{pmatrix}E_1^\rightarrow \\ E_1^\leftarrow\end{pmatrix}.
\end{equation}
For instance, if light (with a wavelength $\lambda_0$ in vacuum) propagates a distance $d$ in a homogeneous dielectric having a refractive index $n$, the transfer matrix is
\begin{equation}
M_d=
\begin{pmatrix}
\exp{-\frac{2\ii\pi nd}{\lambda_0}}& 0\\
0& \exp{\frac{2\ii\pi nd}{\lambda_0}}
\end{pmatrix}.
\end{equation}
If the system is an interface between two dielectrics $n_1$ and $n_2$ perpendicular to the wave vector, the Fresnel formulae allow us to write the transfer matrix of the interface~\cite{Berry1997}
\begin{equation}
N_{n_1\rightarrow n_2}=\frac{1}{2n_2}
\begin{pmatrix}
n_2+n_1& n_2-n_1\\
n_2-n_1& n_2+n_1
\end{pmatrix}.
\end{equation}
The transfer matrix of a dielectric slab of refractive index $n$ and thickness $d$ placed in air ($n_0$) is then written as the matrix product $N_{n\rightarrow n_0}\times M_{d} \times N_{n_0\rightarrow n}$.

Faraday rotation, which is a kind of circular birefringence, imposes to split the field in the two circular polarization components $E^{\rightarrow/\leftarrow}=\begin{pmatrix}E^{\mathrm{R}\rightarrow/\leftarrow},E^{\mathrm{L}\rightarrow/\leftarrow}\end{pmatrix}$.
In a Faraday active material, these components are related to each other by a $4\times 4$ matrix $M_\mathrm{FR}$,
\begin{equation}
\begin{pmatrix}
E_{d}^{\mathrm{R}\rightarrow}\\
E_{d}^{\mathrm{L}\rightarrow}\\
E_{d}^{\mathrm{R}\leftarrow}\\
E_{d}^{\mathrm{L}\leftarrow}\\
\end{pmatrix} = M_\mathrm{FR} (d,\Delta n)
\begin{pmatrix}
E_{0}^{\mathrm{R}\rightarrow}\\
E_{0}^{\mathrm{L}\rightarrow}\\
E_{0}^{\mathrm{R}\leftarrow}\\
E_{0}^{\mathrm{L}\leftarrow}\\
\end{pmatrix},
\end{equation}
\begin{equation}
M_{\mathrm{FR}}=
\begin{pmatrix}
 \exp{-\ii n_\mathrm{+}\varphi}& 0& 0& 0\\
 0& \exp{-\ii n_\mathrm{-}\varphi}& 0& 0\\
 0& 0& \exp{\ii n_\mathrm{+}\varphi}& 0\\
 0& 0& 0& \exp{\ii n_\mathrm{-}\varphi}
\end{pmatrix},
\end{equation}
with $n_\pm=n\pm \Delta n$ and $\varphi=\frac{2\pi d}{\lambda_0}$.
Here $2\Delta n$ is the refractive index difference for right and left circular polarized waves, and is related to the Verdet constant $V$ of the Faraday active material and to the applied magnetic field $B$ via $\Delta n=\frac{\lambda_0 BV}{2\pi}$.
Note that a very similar matrix $M_{\mathrm{OA}}$ can be written for optical activity (which does not break reciprocity),
\begin{equation}
M_{\mathrm{OA}}=
\begin{pmatrix}
 \exp{-\ii n_{+}\varphi}& 0& 0& 0\\
 0& \exp{-\ii n_{-}\varphi}& 0& 0\\
 0& 0& \exp{\ii n_{-}\varphi}& 0\\
 0& 0& 0& \exp{\ii n_{+}\varphi}
\end{pmatrix}.
\label{eq:MFR}
\end{equation}

Using the Fresnel formulae between a dielectric and a Faraday active material (see Appendix~\ref{sec:fresnel}, Table~\ref{tab:fresnel}), we can write in this (circular) basis the transfer matrix of an interface between air (of refractive index $n_0$) and a Faraday active material,
\begin{equation}
N_{n_0\rightarrow n_{+},n_{-}}=
\frac{1}{2}
\begin{pmatrix}
\frac{n_0+n_{+}}{n_{+}} & 0 & 0 & \frac{n_{+}-n_0}{n_{+}}\\
0 & \frac{n_0+n_{-}}{n_{-}} & \frac{n_{-}-n_0}{n_{-}} & 0\\
0 & \frac{n_{-}-n_0}{n_{-}} & \frac{n_0+n_{-}}{n_{-}} & 0\\
\frac{n_{+}-n_0}{n_{+}} & 0 & 0 & \frac{n_0+n_{+}}{n_{+}}
\end{pmatrix},
\end{equation}
and the transfer matrix of an interface between a Faraday active material and air,
\begin{equation}
N_{n_{+},n_{-}\rightarrow n_0}=
\frac{1}{2n_0}
\begin{pmatrix}
{\scriptstyle n_0+n_{+}} & 0 & 0 & {\scriptstyle n_0-n_{+}}\\
0 & {\scriptstyle n_0+n_{-}} & {\scriptstyle n_0-n_{-}} & 0\\
0 & {\scriptstyle n_0-n_{-}} & {\scriptstyle n_0+n_{-}} & 0\\
{\scriptstyle n_0-n_{+}} & 0 & 0 & {\scriptstyle n_0+n_{+}}
\end{pmatrix}.
\label{eq:MOA}
\end{equation}

In the experiments we worked with a linearly polarized incoming field $E_{0}^\rightarrow=\begin{pmatrix}E_{0}^{x\rightarrow}\\E_{0}^{y\rightarrow}\end{pmatrix}$ such that we start these calculations in the linear basis.
Changing from the circular to the linear basis can be done using the matrices for the basis change $P_\mathrm{circ \rightarrow lin}$ and $P_\mathrm{lin \rightarrow circ}$ (Eqs.~(\ref{eq:circlin}) and (\ref{eq:lincirc})).
Without any loss of generality let us consider an $x$-polarized incident wave.
This implies $E_{0}^{y\rightarrow}=0$.
Having just a light source before the system, the field after the system is connected via the transfer matrix $\mathcal{M}$ to the field before the system,
\begin{align}
\begin{pmatrix}
E_{N}^{x\rightarrow}\\
E_{N}^{y\rightarrow}\\
0\\
0\\
\end{pmatrix}
=
\mathcal{M}
\begin{pmatrix}
E_{0}^{x\rightarrow}\\
0\\
E_{0}^{x\leftarrow}\\
E_{0}^{y\leftarrow}\\
\end{pmatrix},
\label{eq:system}
\end{align}
with $E_{0}^\leftarrow=\begin{pmatrix}E_{0}^{x\leftarrow}\\E_{0}^{y\leftarrow}\end{pmatrix}$ the field reflected by the system.
Note that there is no field in the $-z$ direction ($\leftarrow$) after the system and therefore $E_{N}^{x\leftarrow}=E_{N}^{y\leftarrow}=0$.
The overall transfer matrix $\mathcal{M}$ that connects the incoming and the outgoing fields through a stack of $N\geq 2$ Faraday active slides of thicknesses $\left\{d_i\right\}$ separated by $N-1$ air gaps of thicknesses $\left\{e_i\right\}$ in the linear basis is then expressed by
\begin{widetext}
\begin{align}
\mathcal{M}=&P_{\mathrm{circ\rightarrow lin}} \times  
\prod_{i=N}^{2} \left[N_{n_{+},n_{-}\rightarrow n_0} M_{\mathrm{FR}}(d_i,\Delta n) N_{n_0\rightarrow n_{+},n_{-}} M_{0}(e_i)\right] 
\times N_{n_{+},n_{-}\rightarrow n_0} M_{\mathrm{FR}}(d_1,\Delta n) N_{n_0\rightarrow n_{+},n_{-}}
\times  P_{\mathrm{lin\rightarrow circ}},
\label{eq:transferMatrix}
\end{align}
$M_0(e_i)$ being the transfer matrix of an air gap.

Knowing $\mathcal{M}$, the system (\ref{eq:system}) has four equations and four unknowns ($E_{0}^{x\leftarrow}$, $E_{0}^{y\leftarrow}$, $E_{N}^{x\rightarrow}$ and $E_{N}^{y\rightarrow}$) and can therefore be solved analytically.
Writing $\mathcal{M}=\left(m_{ij}\right)_{0\leq i,j < 4}$ using its matrix elements, the solution is:
\begin{align}
\begin{cases}
E_{0}^{x\leftarrow}  &= E_{0}^{x\rightarrow} \cdot \frac{-m_{20}m_{33} + m_{23}m_{30}}{m_{22}m_{33} - m_{23}m_{32}}\\
E_{0}^{y\leftarrow}  &= E_{0}^{x\rightarrow} \cdot \frac{m_{20}m_{32} - m_{22}m_{30}}{m_{22}m_{33} - m_{23}m_{32}}\\
E_{N}^{x\rightarrow} &= E_{0}^{x\rightarrow} \cdot \left(m_{00}+  \frac{- m_{20}(m_{02}m_{33} - m_{03}m_{32}) + m_{30}(m_{02}m_{23} - m_{03}m_{22})}{m_{22}m_{33} - m_{23}m_{32}}\right)\\
E_{N}^{y\rightarrow} &= E_{0}^{x\rightarrow} \cdot \left(m_{10}+  \frac{- m_{20}(m_{12}m_{33} - m_{13}m_{32}) + m_{30}(m_{12}m_{23} - m_{13}m_{22})}{m_{22}m_{33} - m_{23}m_{32}}\right).
\end{cases}
\label{eq:solution}
\end{align}
\end{widetext}
This transfer-matrix method is used to calculate the transmission through $N$ slides with and without Faraday rotation. Different realizations of disorder were realized by randomly varying the thicknesses $d_i$ and $e_i$ of both the plates and the spaces between the plates uniformly over $\Delta d$ and $\Delta e \gg \lambda_0$.
\subsection{Fresnel formulae at the boundary of a circular birefringent medium}
\label{sec:fresnel}
We derive the Fresnel formulae at normal incidence between air (refractive index $n_0$) and a Faraday or optical active (FA or OA) medium described by two refractive indices $n_+$ and $n_-$ (see Table~\ref{tab:refIndex} for the definitions).
\begin{table}[b]
    \centering
    \begin{tabular}{cccc}
        \hline
        \hline
    Propagation direction & Polarization & FA & OA\\
    \hline
    forward  $\rightarrow$ & RHC & $n_+$ & $n_+$ \\
    forward  $\rightarrow$ & LHC & $n_-$ & $n_-$ \\
    backward $\leftarrow$ & RHC & $n_+$ & $n_-$ \\
    backward $\leftarrow$ & LHC & $n_-$ & $n_+$ \\
        \hline
        \hline
    \end{tabular}
    \caption{Refractive indices seen by the forward and backward propagating circular right (RHC) and left (LHC) polarizations.}
    \label{tab:refIndex}
\end{table}

\newcommand{\x}{\hat{\mathbf{x}}}
\newcommand{\y}{\hat{\mathbf{y}}}
\newcommand{\z}{\hat{\mathbf{z}}}
\newcommand{\kk}[1]{\mathbf{k}_{#1}}
$\x$, $\y$ and $\z$ are unit vectors defined in Fig.~\ref{fig:1Dgeometry}: $(\x,\y)$ is the polarization plane, and $\z$ is perpendicular to the interface.
We define the right ($\mathbf{R}$) and left ($\mathbf{L}$) basis vectors for the forward ($\rightarrow$) and backward ($\leftarrow$) propagation directions,
\begin{align}
    \begin{cases}
    \mathbf{R}^\rightarrow=\frac{1}{\sqrt 2} \left(\x + i\y \right)\\
    \mathbf{L}^\rightarrow=\frac{1}{\sqrt 2} \left(\x - i\y \right)\\
    \mathbf{R}^\leftarrow=\frac{1}{\sqrt 2} \left(\x - i\y \right)\\
    \mathbf{L}^\leftarrow=\frac{1}{\sqrt 2} \left(\x + i\y \right).
    \end{cases}
\end{align}

\newcommand{\RR}{\mathrm{R}}
\newcommand{\LL}{\mathrm{L}}
As an example, let us now assume that the incident light is circular right polarized and propagates in the forward direction $\mathbf{E}_i = \mathbf{E}_i^\RR = E_i^\RR \mathbf{R}^\rightarrow$.
We define the reflection $r$ and transmission coefficients $t$ for circular right R and left L by
\begin{align}
    \begin{cases}
    \mathbf{E}_r &= E_i^\RR\left(r_\RR \mathbf{R}^\leftarrow  + r_\LL \mathbf{L}^\leftarrow\right)\\
    \mathbf{E}_t &= E_i^\RR\left(t_\RR \mathbf{R}^\rightarrow + t_\LL \mathbf{L}^\rightarrow\right).
    \end{cases}
\end{align}
We call $\mathbf{E}_t^\RR = E_i^\RR t_\RR \mathbf{R}^\rightarrow$ and $\mathbf{E}_t^\LL = E_i^\RR t_\LL \mathbf{L}^\rightarrow$, and write both continuity equations for the electric and the magnetic fields,
\begin{align}
    \begin{cases}
    \z\times \left(\mathbf{E}_i+\mathbf{E}_r\right) &= \z\times\mathbf{E}_t\\
    \z\times \left(\kk{0} \times \mathbf{E}_i- \kk{0}\times \mathbf{E}_r\right) &= \z\times \left(\kk{+}\times \mathbf{E}_t^{R} + \kk{-}\times \mathbf{E}_t^{L}\right),
    \end{cases}
\end{align}
with $\kk{0}$ the wave vector in the dielectric $n_0$, and $\kk{+/-}$ the wave vectors for the components seeing the $n_{+/-}$ indices (all wave vectors are oriented in the $\rightarrow$ direction).
Solving this system at normal incidence gives
\begin{align}
    \begin{cases}
        r_\RR &= 0\\
        r_\LL &= \frac{n_0-n_+}{n_0+n_+}\\
        t_\RR &= \frac{2n_0}{n_0+n_+}\\
        t_\LL &= 0.
    \end{cases}
\end{align}

This procedure can now be done for all propagation directions, circular polarizations, and for both FA and OA materials.
The resulting Fresnel formulae are summarized in Table~\ref{tab:fresnel}.
\begin{table}[b]
    \centering
    \begin{tabular}{cccccc}
    \hline
    \hline
        Direction & Incident pol. & $r_\RR$ & $r_\LL$ & $t_\RR$ & $t_\LL$ \\
    \hline
        Diel $\rightarrow$ FA, OA & R & 0 & $\frac{n_0-n_+}{n_0+n_+}$ & $\frac{2n_0}{n_0+n_+}$ & 0\\
        Diel $\rightarrow$ FA, OA & L & $\frac{n_0-n_-}{n_0+n_-}$ & 0 & 0 & $\frac{2n_0}{n_0+n_-}$\\
        Diel $\leftarrow$  FA     & R & 0 & $\frac{n_--n_0}{n_0+n_-}$ & $\frac{2n_-}{n_0+n_-}$ & 0\\
        Diel $\leftarrow$  FA     & L & $\frac{n_+-n_0}{n_0+n_+}$ & 0 & 0 & $\frac{2n_+}{n_0+n_+}$\\
        FA $\rightarrow$ Diel     & R & 0 & $\frac{n_+-n_0}{n_0+n_+}$ & $\frac{2n_+}{n_0+n_+}$ & 0\\
        FA $\rightarrow$ Diel     & L & $\frac{n_--n_0}{n_0+n_-}$ & 0 & 0 & $\frac{2n_-}{n_0+n_-}$\\
        FA $\leftarrow$  Diel     & R & 0 & $\frac{n_0-n_-}{n_0+n_-}$ & $\frac{2n_0}{n_0+n_-}$ & 0\\
        FA $\leftarrow$  Diel     & L & $\frac{n_0-n_+}{n_0+n_+}$ & 0 & 0 & $\frac{2n_0}{n_0+n_+}$\\
        Diel $\leftarrow$  OA     & R & 0 & $\frac{n_+-n_0}{n_0+n_-}$ & $\frac{n_++n_-}{n_0+n_-}$ & 0\\
        Diel $\leftarrow$  OA     & L & $\frac{n_--n_0}{n_0+n_+}$ & 0 & 0 & $\frac{n_++n_-}{n_0+n_+}$\\
        OA $\rightarrow$ Diel     & R & 0 & $\frac{n_+-n_0}{n_0+n_-}$ & $\frac{n_++n_-}{n_0+n_-}$ & 0\\
        OA $\rightarrow$ Diel     & L & $\frac{n_--n_0}{n_0+n_+}$ & 0 & 0 & $\frac{n_++n_-}{n_0+n_+}$\\
        OA $\leftarrow$  Diel     & R & 0 & $\frac{n_0-n_+}{n_0+n_+}$ & $\frac{2n_0}{n_0+n_+}$ & 0\\
        OA $\leftarrow$  Diel     & L & $\frac{n_0-n_-}{n_0+n_-}$ & 0 & 0 & $\frac{2n_0}{n_0+n_-}$\\
    \hline
    \hline
    \end{tabular}
    \caption{Fresnel coefficients calculated with the refractive indices convention of Table~\ref{tab:refIndex} for light propagating in the forward $\rightarrow$ or backward $\leftarrow$ direction.}
    \label{tab:fresnel}
\end{table}
\begin{figure*}
    \includegraphics[width=\textwidth]{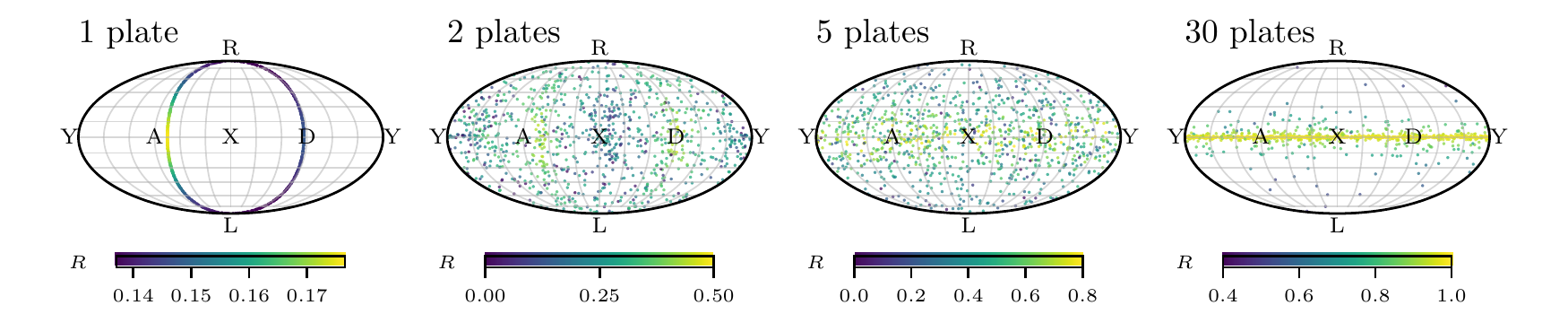}
    \caption{Simulated polarization states in reflection for samples of $N=1, 2, 5$ or 30 plates represented on the Poincaré sphere (Mollweide projection) at $B=18$\,T for the Faraday active plates studied in the paper.
    The color encodes the total reflection ($R=R_x+R_y$).
    X (respectively Y) represents linear polarization on the $x$ (respectively $y$) axis, D and A stand for diagonal and anti-diagonal (linear polarization on the first and second bisectors), and R/L are the circular polarizations (right/left).}
    \label{fig:poincareRefl}
\end{figure*}

\subsection{Transfer matrix and linear/circular polarization}

To include Faraday rotation in these calculations it is useful to switch from a linear basis $\begin{pmatrix}x\\y\end{pmatrix}$ to a circular basis  $\begin{pmatrix}\RR\\\LL\end{pmatrix}$, as Faraday rotation can also be seen as circular birefringence.
For propagation in the $+z$ direction ($\rightarrow$) and respectively for the $-z$ direction ($\leftarrow$), the fields in the circular basis can be expressed as 
$$
\begin{cases}
E_R^+=\frac{1}{\sqrt 2} \left(E_x^+ + i E_y^+\right)\\
E_L^+=\frac{1}{\sqrt 2} \left(E_x^+ - i E_y^+\right)\\
E_R^-=\frac{1}{\sqrt 2} \left(E_x^- - i E_y^-\right)\\
E_L^-=\frac{1}{\sqrt 2} \left(E_x^- + i E_y^-\right)\\ 
\end{cases} \ .
$$
With this, the matrix for the basis change between linear 
$\begin{pmatrix}
E_{x}^+\\
E_{y}^+\\
E_{x}^-\\
E_{y}^-\\
\end{pmatrix}$
and circular 
$\begin{pmatrix}
E_{\mathrm{R}}^+\\
E_{\mathrm{L}}^+\\
E_{\mathrm{R}}^-\\
E_{\mathrm{L}}^-\\
\end{pmatrix}$
polarized fields can be calculated:
\begin{align}
P_\mathrm{circ\rightarrow lin}=\frac{1}{\sqrt 2}\begin{pmatrix}
1& 1& 0& 0\\
i&-i& 0& 0\\
0& 0& 1& 1\\
0& 0&-i& i\end{pmatrix},
\label{eq:circlin}
\end{align}

\begin{align}
P_\mathrm{lin\rightarrow circ}=\frac{1}{\sqrt 2}\begin{pmatrix}
1&-i& 0& 0\\
1& i& 0& 0\\
0& 0& 1& i\\
0& 0& 1&-i\end{pmatrix}.
\label{eq:lincirc}
\end{align}

\section{Magneto-optical spreading of polarization}
\label{sec:polSpreading}

\subsection{Transfer-matrix simulations in reflection}
\label{sec:polSpreadingNum}
\subsubsection{Polarization states}
\label{sec:polStates}

Figure~\ref{fig:poincareRefl} shows the simulated polarization states in reflection for different realizations of disorder for samples consisting of $N=1, 2, 5$ or 30 plates when an external magnetic field of 18\,T is applied on a Faraday active sample made of SF57.
The incident light is linearly polarized along the $x$ axis.
Without magnetic field, the reflected light stays linearly polarized along the $x$ axis.
With magnetic field (18\,T, V=31\,rad/Tm), one can observe that the polarization states get indeed randomized over the whole Poincaré sphere for intermediate plate number, before being attracted on the equator of the sphere.
This means that for large plate numbers, the reflected light is always linearly polarized, but with a random polarization direction.
For optical activity no such change of polarization changes is observed: all the points are exactly on the X point (figure not shown), meaning that for any sample at any optical activity strength, the reflected photons are always linearly polarized in the incident polarization direction.

\subsubsection{Refection coefficients in the case of Faraday rotation and optical activity}
\label{sec:reflectionCoeffFROA}
\begin{figure}
	\includegraphics[width=\columnwidth]{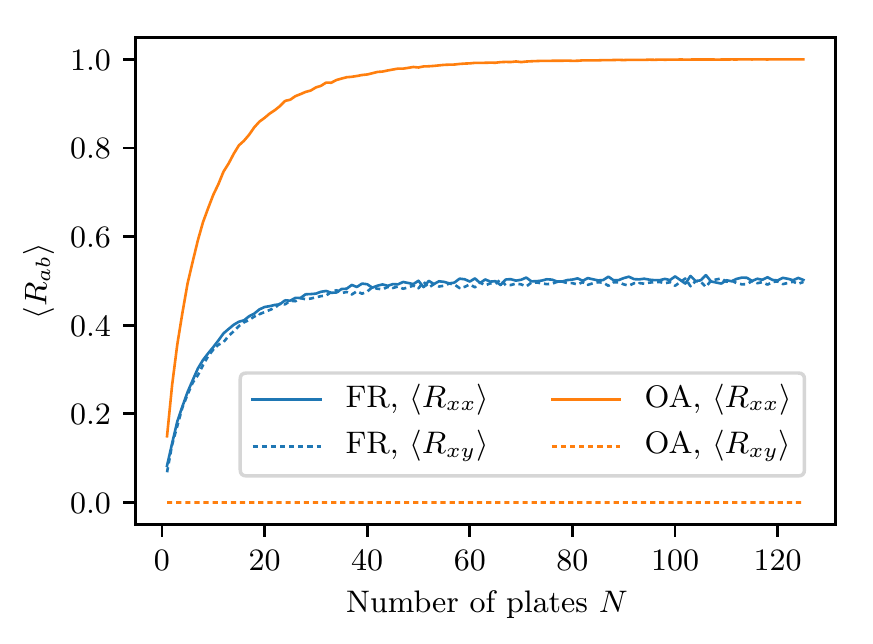}
    \caption{Simulated reflection coefficients in the case of the same circular birefringence plotted against the sample length, for Faraday rotation and for optical activity for the same rotation angle of 48$^\circ$ for a single passage through a single plate. This corresponds to a circular birefringence $\Delta n=4.72\times 10^{-5}$ for FR in SF57 at $B=18T$\ and at the average thickness of the plates.}
    \label{fig:bsc}
\end{figure}
We plot in Fig.~\ref{fig:bsc} the ensemble average over disorder of the reflection coefficients $R_{xx}$ and $R_{xy}$ as a function of the length of the sample, for an external magnetic field of 18 Tesla and compare it to the case were $M_\mathrm{FR}$ (Eq.~(\ref{eq:MFR})) is replaced by $M_\mathrm{OA}$ (Eq.~(\ref{eq:MOA})).
Without circular birefringence, $\left<R_{xy}\right>=0$ because the two channels are uncoupled.
When circular birefringence is turned on, for a large enough sample, optical activity implies $\left<R_{xx}\right>=1$ and $\left<R_{xy}\right>=0$ (the reflected photons leave the sample with the same polarization as the incident light), while Faraday rotation implies $\left<R_{xx}\right>=\left<R_{xy}\right>=0.5$ (the polarization of the reflected photons is randomized).
These observations illustrate the fact that Faraday rotation breaks reciprocity, while optical activity does not.

\subsection{Experimental observation in transmission}
\label{sec:polSpreadingExp}
\begin{figure}
    \includegraphics[width=\columnwidth]{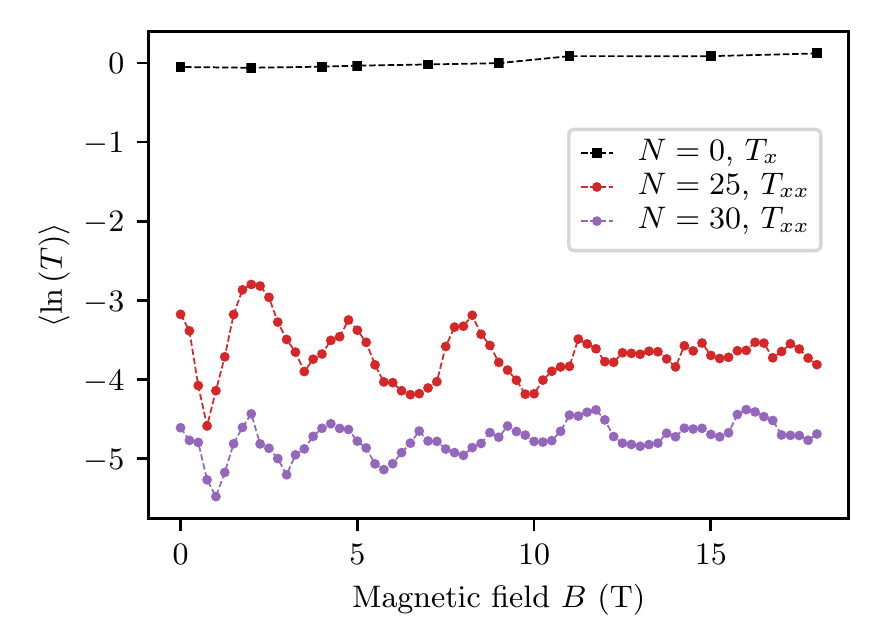}
    \caption{Experimental data of the mean logarithmic relative transmission as a function of the magnetic field $B$ for $N=0$, 25 and 30 plates.
    }
    \label{fig:B_StaticTransExp}
\end{figure}

The spreading of the polarization states in transmission is discussed in Fig.~\ref{fig:poincareTrans} for the transfer-matrix simulations.
On the experimental side, the strong average Faraday rotation and its progressive spreading (depolarization) is also directly seen in the polarized average transmission. Figure~\ref{fig:B_StaticTransExp} shows $\langle \ln T \rangle$ plotted as a function of the external magnetic field $B$ for different fixed numbers of plates.
The $N=0$ measurement (black points) is very stable and used as reference.
The oscillations of $\langle \ln T_{xx} \rangle$ due to the non reflected part of the light beam are again observed (red and violet data in Fig.~\ref{fig:B_StaticTransExp}), and they smear out for large field when the distribution of the polarization states has become almost symmetric around the polar axis of the Poincar\'e sphere.
The pseudo-period of the oscillations decreases with the number of plates as there is more average Faraday rotation as well as higher variance of FR in longer samples (compare $N=25$ plates (red) and $N=30$ plates (violet) data of Fig.~\ref{fig:B_StaticTransExp}).

Note that a field-dependent oscillation behavior of $\left<\ln T\right>$ such as seen in Fig.~\ref{fig:B_StaticTransExp} is not found in genuine-Q1D (i.e., with reflecting walls~\cite{Chabanov2000}) or 3D multiple-scattering situations where polarization memory is lost on the scale of the transport mean free path \ls  and, as a consequence, the overall average FR remains zero in transmission~\cite{Lenke2000}. The oscillations highlight the fact that light propagation in 1D or Q1D stacks of slides is essentially parallel or antiparallel to $B$, while it is random in situations of diffuse multiple-scattering (when \ls is smaller that the transverse scale of confinement).

\section{Data analysis}
\label{sec:dataAnalysis}

\begin{figure}
    \includegraphics[width=\columnwidth]{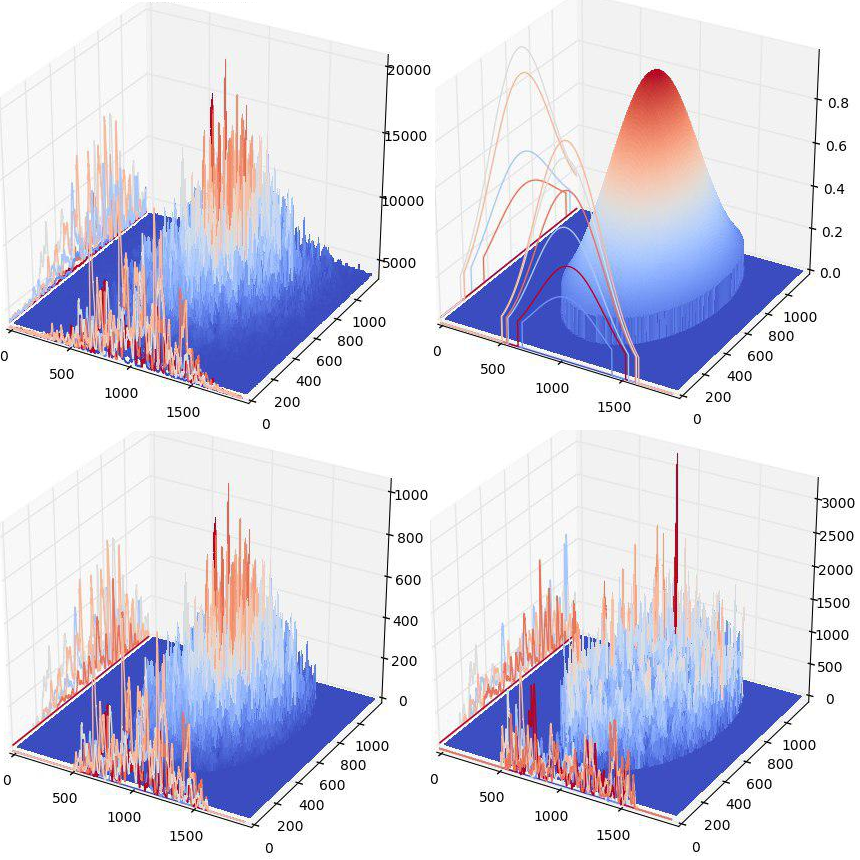}
    \caption{3D image of a measured speckle pattern before image processing (upper left). The intensity is plotted on the $z$ axis while $x$ and $y$ are the camera pixel values. A Gaussian was fitted with an offset and an elliptical region of interest was defined (upper right). The offset was then subtracted from the image (lower left) and the image was weighted by the Gaussian to remove the intensity differences caused by the incident Gaussian laser beam.}
    \label{fig:1DSpeckleimage3D}
\end{figure}
The measured images (see Fig.~\ref{fig:pic-plates}) were processed to obtain the mean transmission $\left<T\right>$ (and/or the mean of the logarithm of the transmission $\langle \ln T \rangle$) and the variance of the relative transmitted intensity $s=T/\left<T\right>$.
Figure~\ref{fig:1DSpeckleimage3D} shows the different steps of the image processing for one measured image of a transmission speckle.

The intensity variations caused by the incident Gaussian beam need to be removed to obtain the correct intensity statistics. A 3D plot of an original image is shown in the upper left. A speckle pattern overlaying the Gaussian beam can be observed. The intensity data were fitted by a 2D Gaussian with widths $\sigma_x$ and $\sigma_y$ and an offset $b$. The offset is necessary to account for noise and was subtracted from the images (lower left). Moreover, an elliptical region of interest (ROI) was chosen with widths 2$\sigma_x$ and 2$\sigma_y$ respectively (see cutoff in the upper right image). The elliptical form is necessary since for a large number of plates the transmitted beam becomes elliptical due to non-perfect alignment of the sample regarding the beam direction. This ellipse rotates with the rotating sample such that the ROI was further narrowed by taking into account only the overlap of the ellipses of all 300 images. The ROI was divided by the Gaussian fit to obtain the final images as shown in the lower right in Fig.~\ref{fig:1DSpeckleimage3D}.
For an increasing number of glass slides increasing laser powers were used as the transmission decreases with the number of plates. Therefore, the transmitted images were normalized by the incident laser power.


\bibliography{Bibliographie}

\end{document}